\documentstyle{article}
\def \be{\begin{equation}}
\def \ee{\end{equation}}
\def \lb{\label}
\def \ba{\begin{array}{l}}
\def \ea{\end{array}}

\def \t{\tau}
\def \a{\alpha}

\def \d{\delta}
\def \D{\Delta}
\def \e{\epsilon}
\def \f{\phi}

\def \lm{\lambda}
\def \n{\nabla}
\def \p{\varphi}
\def \tl{\tilde}
\def \ol{\overline}

\def \l{\left}

\def \fr{\frac}
\def \R{R_{c}}
\def \T{T_{c}}

\def \I{\int d^{D}x}

\def \2{\frac{1}{2}}
\def \4{\frac{1}{4}}

\newcommand{\sectio}[1]{\section{#1}\setcounter{equation}{0}}

\begin{document}

\begin{large}

\vspace{2cm}

\begin{center}

{\Large \bf Replica Symmetry Breaking in the Critical Behaviour
of the Random Ferromagnet}
\vskip .4in
Viktor Dotsenko,$^{1,2}$ A. B. Harris,$^{1,3}$ David Sherrington,$^1$
and R. B. Stinchcombe$^1$
\vskip .2in
1.  Theoretical Physics, Oxford University, \\
1 Keble Rd.,  Oxford OX1 3NP, UK \\
\vskip .2in
2.  Landau Institute for Theoretical Physics,\\
Russian Academy of Sciences, Moscow, Russia\\
\vskip .2in
3.  Department of Physics, University of Pennsylvania, \\
Philadelphia, PA 19104-6396, USA \\
{}~\\
{}~\\
\today

\end{center}
\vskip 1in

\begin{abstract}
We study the critical properties of the weakly disordered
$p$-component random Heisenberg ferromagnet. It is shown that if
the specific heat critical exponent of the pure system is positive,
the traditional renormalization group (RG) flows at dimensions
$D=4-\e$, which are usually considered as describing the
disorder-induced universal critical behavior, are {\it unstable}
with respect to replica symmetry breaking (RSB) potentials as found in
spin glasses. It is demonstrated that the RG flows involving RSB
potentials lead to fixed points which have the structure
known as the 1 step RSB, and there exists a whole spectrum of such
fixed points.  It is argued that spontaneous RSB can occur due to the
interactions of the fluctuating fields with the local non-perturbative
degrees of freedom coming from the multiple local minima solutions of the
mean-field equations.  However, it is not clear whether or not RSB
occurs for infinitesimally weak disorder.  Physical consequences of
these conclusions are discussed.
\end{abstract}
\newpage

\sectio{INTRODUCTION}

\indent
It has been many years since the effects produced by weak quenched
disorder on the critical phenomena were considered to be
qualitatively understood.  In the most general terms the traditional
point of view could be summarized as follows.

If the disorder is weak (e.g. the concentration of impurities
is small), its effect on the critical behavior in the vicinity of
the phase transition point $\T$ remains negligible so long as the
correlation length $\R$ is not too large, i.e. for temperatures
$T$ not too close to $T_{c}$. In this regime the critical behavior
will be essentially the same as in the pure system. However, as
$\t \equiv (T-\T)/\T \to 0$ and $\R (\tau)$ becomes larger than
the average distance between impurities, their influence can
become crucial.

As $\T$ is approached the following change of length scale takes
place.  First, the correlation length of the fluctuations becomes
much larger than the lattice spacing (which we take to be unity),
and the system "forgets" about the lattice.  The only relevant
scale that remains in the
system in this regime is the correlation length $\R(\t)$. Then,
in the close vicinity of the critical point, $\R$ grows and
becomes larger than the average distance between the impurities,
so that the effective concentration of impurities, measured with
respect to the correlation length, becomes large. Such a
situation is reached for an arbitrary small concentration $u$ of
impurities.  The strength of disorder, as scaled by $u$, affects
only the width of the temperature region near $\T$ in which the
effective concentration gets large. If $u \R^{D} \gg 1$, where
$D$ is the spatial dimensionality, one has no grounds, in
general, for believing that the effect of impurities will be
small.

A very simple general criterion has been discovered, the so-called
Harris criterion \cite{harris}, which makes it possible to predict
the effect of impurities qualitatively from only the critical
exponents of the pure system.  According to this criterion the
impurities change the critical behavior only if $\alpha$, the
specific heat exponent of the pure system, is greater than zero
(i. e. the specific heat of the pure system is divergent at the
critical point); more properly, this criterion should be stated
as $D \nu > 2$, where $\nu$ is the correlation length exponent.)
According to the traditional point of view, when this criterion
is satisfied, a new universal critical behavior, with new critical
exponents, is established sufficiently close to the phase transition
point \cite{new,newnew}.  These new exponents were found to satisfy the
opposite criterion (i. e. $\alpha$ was negative) and were
therefore apparently stable.  In contrast, when $\a < 0$ (the
specific heat is finite), the impurities appear to be irrelevant,
i.e. their presence does not affect the critical behavior.

We now consider this point in more detail. Near the phase transition
point the $D$-dimensional Ising-like systems can be described in
terms of the scalar field Ginsburg-Landau Hamiltonian with a
double-well potential:

\be
\lb{aaaa}
H = \I \Biggl[ \2 (\n\f(x))^{2} + \2 [\t - \d\t(x)] \f^{2}(x) +
 \frac{1}{4}g \f^{4}(x) \Biggr] \ .
\ee
Here the quenched disorder is described by random fluctuations of
the effective transition temperature $\d\t(x)$ whose probability
distribution is taken to be symmetric and Gaussian:

\be
\label{aaab}
P[\d\t] = p_{0} \exp \Biggl( -\frac{1}{4u}\I (\d\t(x))^{2} \Biggr) \ ,
\ee
where $u \ll 1$ is the small parameter which describes the disorder,
and $p_{0}$ is the normalization constant.  In Eq. (\ref {aaaa})
$\tau \sim (T-T_c)$ and for notational simplicity, we define the sign
of $\delta \tau (x)$ so that positive fluctuations lead to
locally ordered regions, whose effects are the object of our study.

Configurations of the fields $\f(x)$ which correspond to local
minima in $H$ satisfy the saddle-point equation:
\be
\lb{aaac}
-\D\f(x) + \t\f(x) + g\f^{3}(x) = \d\t(x)\f(x) \ .
\ee
Such localized solutions exist in regions of space where
$\tau - \delta \tau (x)$ assumes negative values.
Clearly, the solutions of Eq. (\ref{aaac})
depend on a particular configuration of the
function $\d\t(x)$ being inhomogeneous.
Let us estimate under which conditions the quenched
fluctuations of the effective transition temperature
are the dominant
factor for the local minima field configurations.

Let us consider a large region $\Omega_{L}$ of a linear size $L >> 1$.
The spatially average value of the function $\d\t(x)$ in this region
could be defined as follows:
\be
\lb{aaad}
\d\t(\Omega_{L}) = \frac{1}{L^{D}} \int_{x\in\Omega_{L}} d^{D}x
\d\t(x) \ .
\ee
Correspondingly, for the characteristic value of the temperature
fluctuations (averaged over realizations) in this region one gets:
\be
\lb{aaae}
\d\t_{L} = \sqrt{\ol{\d\t^{2}(\Omega_{L})}} = \sqrt{2u} L^{-D/2} \ .
\ee
Then, the average value of the order parameter $\f(\Omega_{L})$ in this
region can be estimated from the equation:
\be
\lb{aaaf}
-\t + g\f^2 = \d\t(\Omega_{L}) \ .
\ee
One can easily see that if the value of $\t$ is sufficiently small,
i. e. if
\be
\lb{aaag}
\d\t(\Omega_{L}) >> \t
\ee
then the solutions of Eq.(\ref{aaaf}) are defined only by the value
of the random temperature:
\be
\lb{aaah}
\f(\Omega_{L}) \simeq  \pm(\frac{\d\t(\Omega_{L})}{g})^{1/2} \ .
\ee
Now let us estimate up to which sizes of locally ordered regions
this may occur.  According to Eq.(\ref{aaae}) the condition
$\d\t(\Omega_{L}) >> \t$ yields:
\be
\lb{aaai}
L << \frac{u^{1/D}}{\t^{2/D}} \ .
\ee
On the other hand, the estimation of the order parameter in terms of the
saddle-point equation (\ref{aaaf}) could be correct only at scales
much larger than the correlation length $\R \sim \t^{-\nu}$.
Thus, one has the lower bound for $L$:
\be
\lb{aaaj}
L >> \t^{-\nu} \ .
\ee
Therefore, quenched temperature fluctuations are relevant when
\be
\lb{aaak}
\t^{-\nu}<< \frac{u^{1/D}}{\t^{2/D}}
\ee
or
\be
\lb{aaal}
\t^{2 - \nu D} << u \ .
\ee
According to the scaling relations, one has $2-\nu D = \a$.  Thus
one recovers the Harris criterion: if the specific heat of the
pure system is positive, then in the temperature interval,
\be
\lb{aaam}
\t < \t_{*} \equiv u^{1/\a}
\ee
the disorder becomes relevant.  This argument identifies
$1/ \alpha$ as the cross-over exponent associated with
randomness.\cite{AA}

The above considerations also demonstrate another very
important point: in the disorder dominated region one finds
a {\it macroscopic} number of local minimum solutions of the
saddle-point equation (\ref{aaac}).  Indeed, for a given
realization of the random function $\d\t(x)$ there exists
a macroscopic number of spatial "islands" where
$\tau - \d\t(x)$ is negative (so that the local effective
temperature is below $\T$), and in each of these "islands"
one finds two local minimum configurations of the field:
one which is "up", and another which is "down", as
indicated in Eq.(\ref{aaah}).
These local minimal energy configurations are separated by
finite energy barriers, whose heights become larger
as the size of the "islands" are increased.

Now, if one is interested in the critical properties of the
system, one has to integrate over all local field
configurations up to the scale of the correlation length.
This type of calculation is usually performed using a
Renormalization Group (RG) scheme, which self-consistently
takes into account all the fluctuations of the field on
length scales up to $\R$.

The point, however, is that the traditional RG approach is
only a perturbative theory (albeit a powerful one) in which
one treats the deviations of the field around the ground
state configuration, and it can not take into account
other local minimum configurations which are "beyond barriers".
This problem does not arise in the pure systems, where the
solution of the saddle-point equation is unique. However, in
a situation like that discussed above, when one gets numerous
local minimum configurations separated by finite barriers, the
direct application of the traditional RG scheme may be questioned.

In a systematic approach one would like to integrate in an
RG way over fluctuations around the local minima configurations.
Furthermore, one also has to sum over all these local minima up
to the scale of the correlation length. In view of the fact that
the local minima configurations are defined by the random
quenched function $\d\t(x)$ in an essentially non-local way, the
possibility to implement successfully such a systematic approach
seems rather hopeless.

On the other hand there exists another technique which has been
developed specifically for dealing with systems which exhibit
numerous local minima states.  It is the Parisi Replica Symmetry
Breaking (RSB) scheme which has proved to be essential in the
mean-field theory of spin-glasses (see e.g. \cite{sg}). Recent
studies show that in certain cases the RSB approach can also be
generalized for situations where one has to deal with
fluctuations as well \cite{manifold},\cite{rsb-Marc},
\cite{Korshunov}.

In this paper we are going to study the critical properties
of weakly disordered systems in terms of the RG approach
generalized to take into account the possibility of the RSB
phenomena.  The model we will treat is the ferromagnetic
$O(n)$ or Heisenberg model with weak random interactions.
The idea is that hopefully, like in spin-glasses,
this type of generalized RG scheme self-consistently takes
into account relevant degrees of freedom coming from the
numerous local minima. In particular, the instability of the
traditional Replica Symmetric (RS) fixed points with respect to
RSB indicates that the multiplicity of the local minima can be
relevant for the critical properties in the fluctuation region.

It will be shown in the next Section that, whenever the disorder
appears to be relevant for the critical behavior (in accordance
with the Harris criterion), the usual RS fixed points (which used
to be considered as providing new universal disorder-induced
critical exponents) are unstable with respect to "turning
on" an RSB potential. In the presence of such a potential the RG
flows actually go to another type of fixed point having a
structure known as 1-step RSB. The 1-step RSB structure is
described by one parameter $x_{0}$ ($0 < x_{0} < 1$), which is
the coordinate of the RSB step, and this parameter remains
arbitrary within the framework of the RG scheme. Therefore,
within the framework of the formal RG consideration one finds
a whole line of the fixed points (instead of the unique
fixed point in the RS subspace), and correspondingly one obtains
a whole spectrum of critical exponents.

Formally, the value of the parameter $x_{0}$ at the fixed point
is defined by the "initial conditions" for the RG equations.
This situation is qualitatively different from the traditional
RS one, where the fixed point appears to be universal, and does
not depend on the the details of the starting values of the
parameters of the Hamiltonian.  Here, the existence of such
RSB fixed points indicates that the actual values of the critical
exponents could be {\it non-universal}, being dependent on the
concrete characteristics of the disorder involved.

In Section 3 we discuss the problem of the "initial conditions"
for the RG calculations in more detail. The crucial problem for
the present approach is to understand whether RSB is inherent in
the random bond model or not.  If, for instance, RSB does NOT occur
spontaneously for the weakly random bond Heisenberg model, then
one remains in the replica symmetric subspace of potentials and
the traditional approach\cite{new,newnew} is probably valid.
In contrast,
it might be that disorder is always accompanied by a small amount
of RSB.  In that case, critical exponents, amplitude ratios
and the like, would be determined by the RSB fixed point we find
here.  Based on general physical arguments we propose a mechanism
whereby the multiple local minimum solutions discussed above can
provide RSB interactions between the fluctuating fields. It is
argued that to sum over all the local degrees of freedom, one has
to sum over discrete local minima solutions first, and then one
could initiate the RG integrations over the fluctuating fields.
The point is that the fluctuating
fields themselves are the deviations from the local minimum
configurations, and that is why the summation over these
quenched discrete degrees of freedom (together with the replica
averaging over the quenched disorder) could provide additional
non-trivial interactions among the fluctuating fields.  According
to the arguments presented in this Section, the partition function
which describes these interactions is analogous to that of the random
energy model (REM) of Derrida.\cite{rem}  Using the known
results\cite{rem} for the REM, it then follows that
if the interaction, $V_{\rm L-P}$ between local solutions and
perturbative fluctuations exceeds a critical strength,
replica symmetry will be broken.  Thus, we identify a scenario
for obtaining spontaneous RSB from a random spin model.
However, it remains an open question whether, when the bond
randomness is arbitrarily weak, the interaction $V_{\rm L-P}$
exceeds the critical strength necessary for RSB.  If so, the
analogous REM has the 1-step RSB structure and the value of the
coordinate of the step $x_{0}$ is defined by the concrete
statistical characteristics of the disorder involved.

The remaining problems as well as future perspectives
are discussed in the Conclusions.

\sectio{THE RENORMALIZATION GROUP AND REPLICA SYMMETRY BREAKING}

\indent
We consider the $p$-component $\f^{4}$ theory with quenched random
effective temperature fluctuations, which near the transition point
can be described by the usual Ginzburg-Landau Hamiltonian:
\be
\label{bbba}
\begin{array} {l}
H[\d\t,\f] = \I \Biggl[
\2\sum_{i=1}^{p}(\n\f_{i}(x))^{2} \\
\\
+ \2(\t - \d\t(x))\sum_{i=1}^{p}\f_{i}^{2}(x) +
\4 g\sum_{i,j=1}^{p}\f_{i}^{2}(x)\f_{j}^{2}(x) \Biggr] \ ,
\end{array}
\ee
where the quenched potential $\d\t(x)$ is distributed according
to Eq. (\ref{aaab}).  In terms of the standard replica approach
one has to calculate the following replica partition function:
\be
\label{bbbc}
\begin{array}{l}
Z_{n} = \ol{(\int D\f_{i}(x) \exp\{-H[\d\t,\f]\})^{n}} \equiv
\int D \delta \tau P[ \delta \tau ]
(\int D\f_{i}(x) \exp\{-H[\d\t,\f]\})^{n}
\\
\\
= \int D\d\t(x)\int D\f_{i}^{a}(x)
\exp \Biggl[ -\frac{1}{4u}\I [\d\t(x)]^{2} \\
\\
- \I [\2\sum_{i=1}^{p}\sum_{a=1}^{n}[\n\f_{i}^{a}(x)]^{2} +
\2[\t - \d\t(x)]\sum_{i=1}^{p}\sum_{a=1}^{n}[\f_{i}^{a}(x)]^{2} \\
\\
+ \4 g\sum_{i,j=1}^{p}\sum_{a=1}^{n}[\f_{i}^{a}(x)]^{2}[\f_{j}^{a}(x)]^{2}]
\Biggr] \ ,
\end{array}
\ee
where the superscript $a$ labels the replicas. (Here and in what
follows all irrelevant pre-exponential factors are omitted.)
After Gaussian integration over $\d\t(x)$ one gets:
\be
\label{bbbd}
\begin{array}{l}
Z_{n} =
\int D\f_{i}^{a}(x) \exp \Biggl[ - \I \Biggl(
\2\sum_{i=1}^{p}\sum_{a=1}^{n}[\n\f_{i}^{a}(x)]^{2}
+ \2\t\sum_{i=1}^{p}\sum_{a=1}^{n}[\f_{i}^{a}(x)]^{2} \\
\\
+ \4 \sum_{i,j=1}^{p}\sum_{a,b=1}^{n} g_{ab}
[\f_{i}^{a}(x)]^{2}[\f_{j}^{b}(x)]^{2} \Biggr) \Biggr] \ ,
\end{array}
\ee
where
\be
\lb{bbbe}
g_{ab} = g\d_{ab} - u \ .
\ee

To study the critical properties of this system we are going
to apply the standard RG procedure developed for dimensions
$D = 4 - \e$, where $\e \ll 1$.  Along the lines of the usual
rescaling scheme (see e.g. \cite{rg}) one gets the following
RG equations for the interaction parameters $g_{ab}$:

\be
\lb{bbbf}
\fr{d g_{ab}}{d\xi} = \e g_{ab} - \fr{1}{8\pi^{2}}
(4g_{ab}^{2} + 4g_{aa}g_{ab} + p \sum_{c=1}^{n} g_{ac} g_{cb}) \ ,
\ee
where $\xi$ is the standard rescaling parameter.

If one takes the matrix $g_{ab}$ to be replica symmetric,
as in the starting form of Eq. (\ref{bbbe}), then one would
recover the usual RG equations
for the parameters $g$ and $u$, and eventually one would obtain
the well known results for the fixed points and the critical
exponents \cite{new,newnew}.  Here, however, we are going to
study the
stability of the RG flows with respect to RSB in the matrix
$g_{ab}$. We leave the question as to how perturbations out of
the RS subspace could arise until the next Section, and formally
consider the RG eqs.(\ref{bbbf}) assuming that the matrix
$g_{ab}$ has a Parisi RSB structure (which also include the
RS structure as a special case).

According to the standard technique of the Parisi RSB algebra
(see e.g. \cite{sg},\cite{intro}), in the limit $n\to 0$ the
matrix $g_{ab}$ is parametrized in terms of its diagonal
element $\tl g$ and the off-diagonal {\it function} $g(x)$
defined in the interval $0<x<1$:

\be
\lb{bbbg}
g_{ab} \to (\tl g; g(x))
\ee
The RS situation corresponds to the case $g(x) = const$
independent of $x$.  All the operations with the matrices in
this algebra can be
performed according to the following simple rules:

\be
\lb{bbbh}
g_{ab}^{k} \to (\tl g^{k}; g^{k}(x))
\ee
and

\be
\lb{bbbi}
(\hat g^{2} )_{ab} \equiv  \sum_{c=1}^{n} g_{ac} g_{cb} \to (\tl c; c(x))
\ee
where

\be
\lb{bbbj}
\ba
\tl c = \tl g^{2} - \int_{0}^{1} dx g^{2}(x)\\
\\
c(x) = 2(\tl g - \int_{0}^{1} dy g(y) ) g(x) -
\int_{0}^{x} dy [g(x) - g(y)]^{2}
\ea
\ee

Using the above rules, and redefining (just to simplify formulae):
$g_{ab} \to 8\pi^{2} \e g_{ab}$ and $\xi \to \frac{1}{\e} \xi$,
from the eqs.(\ref{bbbf}) one gets:

\be
\lb{bbbk}
\ba
\fr{d}{d\xi}g(x;\xi) = g(x;\xi) - 4g^{2}(x;\xi) - (4+2p)g(x;\xi)\tl{g}(\xi) +\\
\\
+ 2p g(x;\xi) \int_{0}^{1} dy g(y;\xi)
+ p \int_{0}^{x} dy [g(x;\xi)-g(y;\xi)]^{2}\\
\\
\end{array}
\end{equation}
\begin{equation}
\fr{d}{d\xi}\tl{g}(\xi) = \tl{g}(\xi) - (8+p)\tl{g}^{2}(\xi)
+ p\int_{0}^{1} dy g^{2}(y;\xi)
\ee
The usual RS equations are recovered if one takes $g(x;\xi) \equiv g(\xi)$
independent of $x$.

Then, for the fixed point, $\fr{d}{d\xi}g(x) = 0 $,
$\fr{d}{d\xi}\tl{g} = 0$ one obtains the following equations:

\be
\lb{bbb1}
\ba
g(x) - 4g^{2}(x) - (4+2p)g(x)\tl{g} +\\
\\
+ 2p g(x) \int_{0}^{1} dy g(y)
+ p \int_{0}^{x} dy [g(x)-g(y)]^{2} = 0\\
\end{array}
\end{equation}
\begin{equation}
\lb{bbb2}
\tl{g} - (8+p)\tl{g}^{2} + p\int_{0}^{1} dy g^{2}(y) = 0
\end{equation}

It is interesting to note that the structure of Eq.
(\ref{bbb1}) is similar to that for the Parisi
function $q(x)$ near $\T$ in the SK model of spin-glasses (SG).
Although the cubic terms in the "order parameter" function
$g(x)$ are not included here, they could be calculated in the
next order of the loop expansion of the RG.  The most essential
difference from the situation in the SK spin-glasses is the
presence of the term $g^{2}(x)$ in Eq. (\ref{bbb1})
(there is no such term in the SK model).  This is the
typical term which is known to produce 1step RSB in other
SG-like systems \cite{?}, and in this case the higher order
terms produce no qualitative change in the results.

 From the Eqs. (\ref{bbb1}) and (\ref{bbb2})
one can easily find out what should
be the structure of the function $g(x)$ at the fixed point.
Taking the derivative over $x$ twice, one gets, from Eq.
(\ref{bbb1}): $g'(x) = 0$. This means that either the
function $g(x)$ is constant (which is the RS situation), or it
has the step-like structure.

Let us consider the simplest ansatz, which is the 1step RSB
(actually, a bit later we are going to argue, that there are
no many-step fixed points).  Thus we assume that

\be
\lb{bbbm}
g(x) = \l\{ \begin{array}{ll}
g_{0}        & \mbox{for $0 \leq x < x_{0}$}  \\
g_{1}        & \mbox{for $x_{0} < x \leq 1$}
\end{array}
\right.
\ee
where $0 \leq x_{0} \leq 1$ is the coordinate of the step.

Then, from Eqs. (\ref{bbb1}) and (\ref{bbb2}) one gets three
equations for three parameters $\tl g, g_{0}$ and $g_{1}$:

\be
\lb{bbbn}
\ba
(4-2px_{0})g_{0}^{2} - 2p(1-x_{0})g_{1}g_{0} + (4+2p)\tl{g}g_{0} = g_{0}\\
\\
-px_{0}g_{0}^{2} + (4-2p+px_{0})g_{1}^{2} + (4+2p)\tl{g}g_{1} = g_{1}\\
\\
-px_{0}g_{0}^{2} - p(1-x_{0})g_{1}^{2} + (8+p)\tl{g}^{2} = \tl{g}
\ea
\ee
Note again, that for $g_{0} \equiv g_{1}$ one obtains the usual RS
fixed point equations, and the parameter $x_{0}$ drops out from the
equations.

Equations (\ref{bbbn}) have several solutions. Among them there are
two fixed points which are the usual RS ones \cite{new,newnew}:

\be
\lb{bbbo}
g_{0} = g_{1} = 0; \; \; \; \tl{g} = \fr{1}{8+p}
\ee
and

\be
\lb{bbbp}
g_{0} = g_{1} = -\fr{4-p}{16(p-1)}; \; \; \; \tl{g} = \fr{p}{16(p-1)}
\ee

The first of these corresponds to the pure system, and is stable
when the disorder is irrelevant according to the Harris criterion.
The disorder-induced fixed point (\ref{bbbp}) is usually
considered to be the one which describes the new universal critical
behaviour in systems with impurities. This fixed point has
been shown to be stable (with respect to the RS deviations!)
for $p < 4$, which is consistent with the Harris criterion
since the specific heat critical exponent associated with
this fixed point is negative and that associated with the
pure system fixed point is positive.
(For $p=1$ this fixed point involves an expansion in powers of
$( \epsilon )^{1/2}$ \cite{newnew}. This structure is only revealed
within a two-loop approximation and was therefore not located
in the early work \cite{new}.)

However, besides these two RS fixed points there exist the
following two non-trivial 1step solutions with $g_{0} \not= g_{1}$:

\be
\lb{bbbq}
g_{0} = -\fr{4-px_{0}}{16(px_{0}-1)}; \; \; \;
g_{1} = \tl{g} = \fr{px_{0}}{16(px_{0}-1)}
\ee
and

\be
\lb{bbbr}
\ba
g_{0} = 0 ; \; \; \;
g_{1} = -\fr{4-p}{16(p-1) - px_{0}(8+p)}\\
\\
\tl{g} = \fr{p(1-x_{0})}{16(p-1) - px_{0}(8+p)}
\ea
\ee

The stability analysis for the above fixed points with respect
to general RSB deviations would appear to be technically a rather
difficult problem, but one can easily check the stability
within the considered 1step RSB subspace. The calculations,
although a bit cumbersome, are straightforward.  One has to
linearize the "dynamical" equations (\ref{bbbk}) near a given
fixed point, and then one has to find the eigenvalues. The fixed
point is stable only if all three eigenvalues are negative. We
omit this purely technical analysis and report the results only.
The pure system fixed point (\ref{bbbo}) appears to stable
(in accordance with the Harris criterion) for $p > 4$, when the
pure system specific heat critical exponent is negative.

The traditional RS fixed point (\ref{bbbp}) appears to be
{\it always unstable}. The three eigenvalues of the linearized
equations are:
$\lm_{1} = -1/2, \; \; \lm_{2} = - \fr{(4-p)}{8(p-1)} $ and
$\lm_{3} = + \fr{(4-p)}{8(p-1)} $. The instability appears just
due to the possibility of creating a "step".

The first 1step RSB fixed point (\ref{bbbq}) appears to be stable
in the region of parameters: $1 < px_{0} < 4$.

Finally, the second 1step RSB fixed point (\ref{bbbr}) is stable for:

\be
\lb{bbbs}
\ba
1 < p < 4 ;\\
\\
0 < x_{0} < x_{c}(p) \equiv \fr{16(p-1)}{p(8+p)}
\ea
\ee
In particular, $x_{c}(p=2) = 4/5$ and $x_{c}(p=3) = 32/33$
(while $x_{c}(p=4) = 1$).

Note, that in addition to the fixed points listed above there exist several
other 1step RSB solutions of the eqs.(\ref{bbbn}) which we do not reproduce
here because they are always unstable.

Actually the "physical" fixed point is that in Eq. (\ref{bbbr})
(with the stability conditions (\ref{bbbs})), and not that in the
Eq. (\ref{bbbq}), which has to be considered as "unphysical".
The point is that according to the arguments which are due to be
presented in the next Section, whenever the RSB perturbation comes
into play, it always requires (according to its physical origin)
that $g_{1} < g_{0}$.  On the other hand, simple numerical solution
of the evolution Eqs. (\ref{bbbk}) clearly demonstrates that
if the initial values of the parameters are bounded such that
$g_{1} < g_{0}$, then, whatever the actual starting values of
the parameters are, one always ends up at the fixed point
(\ref{bbbr}) and not the one (\ref{bbbq}).  In the opposite
case, $g_{1} > g_{0}$, the RG trajectories always go to the
fixed point (\ref{bbbq}), but this situation must be considered as
"unphysical".

As for the possibility to have many-step solutions of the fixed point
Eqs. (\ref{bbb1}) and (\ref{bbb2}), the generalization of the
above-considered approach
is straightforward, although technically it is much more
cumbersome. On the other hand, the direct numerical observations of
the RG trajectories for the case of several steps in the initial
function $g(x)$ is still rather simple. Such an analysis convincingly
demonstrates that whatever the starting conditions are,
one always ends up in the 1step fixed point (\ref{bbbr}), where
the value of the parameter $x_{0}$ is the coordinate of the
"right-most" step in the initial configuration of the function
$g(x)$.

It should be stressed here that, within the present pure RG
considerations, the value of the parameter $x_{0}$ which defines the
fixed points, remains arbitrary. In this sense, one can say that in
the problem under consideration there exists a whole line of fixed
points instead of the unique universal one.  Formally, the value of
the parameter $x_{0}$ which characterizes a given fixed point is
defined by the initial conditions of the RG equations, and in this
sense one could argue that the critical behavior in systems with
such disorder is not universal.

Another painful problem is to elucidate what is going
on if the value of the parameter $x_{0}$ happens to be in the
"instability region" $x_{c}(p) < x_{0} < 1$. Formally, in this case
the RG trajectories go to infinity, and it means that one has to take
into account the next order RG terms, which
hopefully could stabilize the situation, just as they
do for the Ising ($p=1$) case.\cite{newnew}
We leave these problems for future analysis.

Consider now what the consequences of the existence of the above
1step RSB fixed point are for the critical exponents.  The RG
equations for the mass term of the Hamiltonian in the
Eq.(\ref{bbbd}) are:

\be
\lb{bbbss}
\fr{d}{d\xi}\t_{ab} = 2\t_{ab} -
\e [2\t_{ab}g_{ab} + p\d_{ab}\sum_{c=1}^{n}\t_{cc}g_{cb} ]
\ee
This is a general (one-loop) RG equation for an arbitrary mass matrix
$\t_{ab}$. In our case the initial matrix is diagonal:
$\t_{ab}(\xi = 0) = \t_{0} \d_{ab}$, and it remains diagonal
(in the higher orders of the RG as well) whatever RSB takes place in the
interaction matrix $g_{ab}$. This is just a general consequence of the
absence of fields which would break the symmetry $\f \to -\f$.
At the 1step RSB fixed point (\ref{bbbr}) for the rescaling of
the (diagonal) mass term one gets:

\be
\lb{bbbt}
\ba
\fr{d \ln\t}{d\xi} = 2 - \e [(2+p)\tl{g}
- p(1-x_{0})g_{1} - px_{0}g_{0}] = \\
\\
\\
= 2 - \e \fr{6p(1-x_{0})}{16(p-1)-px_{0}(p+8)} \equiv \D_{\t}
\ea
\ee
According to the standard scaling relations for the critical exponent
of the correlation length one finds:

\be
\lb{bbbu}
\nu(x_{0}) = \fr{1}{\D_{\t}} = \2 + \2\e\fr{3p(1-x_{0})}{16(p-1)-px_{0}(p+8)}
\ee
Correspondingly, for the critical exponent of the specific heat:
$\a = 2 - (4-\e)\nu$, one obtains:

\be
\lb{bbbv}
\a(x_{0}) = -\2\e\fr{(4-p)(4-px_{0})}{16(p-1)-px_{0}(p+8)}
\ee

Thus, depending on the value of the parameter $x_{0}$ one finds
a whole {\it spectrum} of the critical exponents. In particular,
the possible values of the specific heat critical exponent appear to
be in the following band:

\be
\lb{bbbw}
-\infty < \a(x_{0}) < - \e\fr{(4-p)}{8(p-1)}
\ee
The upper bound for $\a(x_{0})$ is achieved in the RS limit $x_{0} \to 0$,
and it coincides with the usual RS result \cite{new}. On the other hand,
as $x_{0}$ tends to the "border of stability" $x_{c}(p)$ of the
1step RSB fixed point, formally the specific heat critical
exponent tends to minus infinity.

Note that, as usual, to obtain the leading fluctuation correction to
the critical exponent of the correlation functions (usually called
$\eta$) one has to study the RG fixed points
in the next order ($\sim \e^{2}$) approximation.

\sectio{POSSIBLE SCENARIO FOR SPONTANEOUS RSB}

\indent
In this Section we will present qualitative arguments showing
how RSB perturbations could be spontaneously generated in the
random bond model.  For simplicity we set $p=1$.
We start by considering the partition function for a fixed
configuration of $\delta \tau (x)$:

\be
\lb{ccca}
Z[\d\t] = \int D\f(x) \exp \{ - H[\f;\d\t]\} \ ,
\ee
where

\be
\lb{cccb}
H [ \phi ; \delta \tau ] = \I \left( \2 [\n\f(x)]^{2}
+ \frac{1}{2}
 [ \tau - \d\t(x)] \f(x)^2 + \frac{1}{4}g \f(x)^4 \right) \ .
\ee
for a given realization $\d\t(x)$, the saddle-point equation,

\be
\lb{cccc}
-\D\f(x) + ( \tau - \d\t\ (x) \phi(x) + g\f(x)^3 = 0
\ee
has many local minima solutions. We denote such a local solution
by $\psi^{(i)} (x)$ with $i=1,2,\dots N_0$.  If the size $L_0$ of an
"island" where $\d\t(x) > 0$ is not too small, then the value of
$\psi^{(i)} (x)$ in this "island" should be $\sim \pm \sqrt{\d\t(x)/g}$,
where $\delta \tau(x)$ should now be interpreted as the value
of $\delta \tau$ averaged over the region of size $L_0$.  Such
"islands" occur at a certain finite density per unit volume.
Thus the number of such local solutions, $N_0$, is macroscopic:
$N_0 = \kappa V$, where $V$ is the volume of the system and
$\kappa$ is a constant.  An approximate global extremal solution
$\Phi (x)$ is constructed as the union of all these local
solutions without regard for interactions between "islands."
Each local solution can occur with either sign, since we are
dealing with the disordered phase:
\begin{equation}
\label{PSI}
\Phi^{(\alpha)} [x; \delta \tau(x)] = \sum_{i=1}^{\kappa V}
\sigma_i \psi^{(i)} (x) \ ,
\end{equation}
where each $\sigma_i = \pm 1$.
Accordingly, the total number of global solutions must be $2^{\kappa V}$.  We
denote these solutions by
$\Phi^{(\alpha )} [x;\d\t (x) ]$,
where $\alpha = 1, 2,...,K = 2^{\kappa V}$.
(In this type of symbol we later write simply $\delta \tau$ for
$\delta \tau (x)$.)
As we mentioned, it seems unlikely that an integration
over fluctuations around $\phi(x)=0$ will include the
contributions from the configurations of $\phi(x)$ which
are near a $\Phi(x)$, since $\Phi(x)$ is "beyond a
barrier," so to speak.  Therefore, it seems appropriate
to include separately the contributions from small fluctuations
about each of the many $\Phi^{(\alpha )} [x; \delta \tau ]$.
Thus we have to sum over the $K$ global minimum solutions
(non-perturbative degrees of freedom)
$\Phi^{(\alpha )} [x; \delta \tau ]$ and also
to integrate over "smooth" fluctuations $\p(x)$ around them:

\be
\lb{cccd}
\begin{array} {l}
Z[\d\t] = \int D\p(x) \sum_{\a}^{K}
\exp \Biggl(  - H \Biggl[
\Phi^{(\alpha )} [x;\delta \tau ] + \p(x); \d\t
\Biggr] \Biggr) \\
\\
= \int D\p(x) \exp \Biggl( -H[\p;\d\t ] \Biggr)
\times \tl{Z}[\p;\d\t] \ ,
\end{array}
\ee
where

\be
\lb{ccce}
\begin{array} {l}
\tl{Z}[\p;\d\t] = \sum_{\a}^{K}
\exp \Biggl( -H[ \Phi^{(\alpha )}] \Biggr) \\
\\
\times \exp \Biggl( -\I \Biggl[
\frac{3}{2}g\Phi^{( \alpha )} [x; \delta \tau ]^2 \p(x)^2 +
g\Phi^{(\alpha )} [x, \delta \tau ] \p(x)^3 \Biggr] \Biggr) \ .
\end{array}
\ee

Next we carry out the appropriate average over quenched disorder.
To do this, we need to average the $n$th ($n \rightarrow 0$) power
of the partition function.  This is accomplished by introducing
the replicated partition function, $Z_n$, as

\be
\lb{cccf}
Z_{n} = \int D\d\t \int D\p_a \exp \left(
 -\fr{1}{4u}\I [ \d\t(x) ]^{2}
- \sum_{a=1}^{n} H[\varphi_a;\d\t] \right)
\times \tl{Z_{n}}[\p_a;\d\t] \ ,
\ee
where the subscript $a$ is a replica index and

\be
\lb{cccg}
\begin{array}{l}
\tl{Z_{n}}[\p_{a};\d\t] =
\sum_{\a_{1}...\a_{n}}^{K}  \\ \\
\exp \Biggl( -\sum_{a}H[\Phi^{( \alpha_a )} ] - \I \sum_{a}
\Biggl[ \frac{3}{2}g \Phi^{(\alpha_a)} (x)^2 \p_a(x)^2 +
g\Phi^{(\a_a)} (x) \p_a(x)^3 \Biggr] \Biggr) \ ,
\end{array}
\ee
where $\Phi(x)$ stands for $\Phi[x, \delta \tau ]$.

It is obviously hopeless to try to make a systematic evaluation
of this replicated partition function.  The global solutions
$\Phi^{(\alpha)}$ are complicated implicit functions of
$\delta \tau (x)$.  These quantities have fluctuations of two
different types.  In the first instance, they depend on the
stochastic variables $\delta \tau (x)$.  But even when
the $\delta \tau (x)$ are completely fixed, $\Phi^{(\alpha)} (x)$
will depend on $\alpha$ (which labels the possible ways of
constructing the global minimum out of the choices for the
signs $\{\sigma\}$ of the local minima).  A crude way of treating
this situation is to regard the local solutions
$\psi^{(i)} (x)$ as if they were random variables, even
though $\delta \tau (x)$ has been specified.  This randomness,
which one can see is not all that different from that which
exists in a spin glass, is the crucial one which we claim
may lead to RSB.  Accordingly, we no longer bother to keep
track explicitly of the fluctuations in $\Phi^{(\alpha )} (x)$
due to its dependence on $\delta \tau (x)$.  Instead, we
introduce a distribution function for $\psi(x)$. The simplest
and the most natural distribution function is the Gaussian one:

\be
\lb{ccch}
\begin{array} {l}
P[ \{ \psi(x) \} ] \equiv P[ \Phi^{(\alpha)} [x;\delta \tau ]]
= \prod_i \exp \Biggl(
-\fr{1}{2\D}\I [\psi^{(i)}(x)^2 - \psi_0(x)^2 ]^2 \Biggr) \\
\\
= \exp \Biggl( - \frac{1}{2\Delta} \int d^D x
[ \Phi^{(\alpha )} (x)^2 -\Phi_0 (x)^2 ]^2 \Biggr) \ ,
\end{array}
\ee
where $\psi_{0}(x) = \sqrt { \d\t(x)/g}$, $\D$ is a parameter,
which in principle should be defined through the original
parameters $g$ and $u$, and $\Phi_0(x)$ is the value of
$\Phi$ given by Eq. (\ref{PSI}) when $\psi^{(i)}(x)$ is
replaced by $\psi_0$.  The above distribution function exhibits
two Gaussian maxima (with a mean square width equal to $\D$)
around the values $\pm \sqrt{\d\t(x)/g}$ in the "islands",
over which, on average, $\d\t(x) >0$.  The width $\D$ reflects
the distribution in the magnitude of $\psi^{(i)} (x)$
which results from distant "islands" fluctuating between
their "up" and "down" states.

Since the distribution (\ref{ccch}) is symmetric with respect
to the signs of the $\psi^{(i)}$, the term
$\Phi^{\a_a} (x)\p_{a}(x)^3$ in the partition function
(\ref{cccg}) can produce only interactions of the order
$\p^{6}$, which are irrelevant for the critical
properties. Therefore, this term can be safely omitted.
Using the equation (\ref{cccc}) for the energy in a given minimum
one easily gets:

\be
\lb{ccci}
H[\Phi^{(\alpha )}] = -\4 \I \Phi^{(\alpha )} (x)^4 \ .
\ee
Then, for the partition function $\tl{Z_{n}}$ one obtains:

\be
\lb{cccj}
\ba
\tl{Z_{n}} \simeq \prod_{\a} \Biggl[ \int D \Phi^{(\alpha )} (x)
\exp \Biggl( -\fr{1}{2\D} \I
[\Phi^{(\alpha )} (x)^2 - \Phi_0 (x)^2 ]^{2} \Biggr) \Biggr] \\
\\
\times\sum_{\a_{1}...\a_{n}}^{K}
\exp \Biggl( \4 g\I \sum_{a}
\Phi^{\a_a}(x)^4-\fr{3}{2}g \I
\sum_{a}  \Phi^{(\a_a)}(x)^2 \p_{a}(x)^2 \Biggr) \ .
\ea
\ee
The non-trivial RSB effects we are looking for come from the
integration in the vicinity of the points
$\psi_i(x)=\pm\psi_0(x)$.  Assuming that the parameter
$\D$ is small enough, one can redefine
$\Phi^{\a}(x)^2 = \Phi_0(x)^2 + z_\alpha (x)$.

Then, for the part of the partition function $\tl{Z_{n}}$ which
contains the integration over $z_{\a}(x)$ one obtains:

\be
\lb{ccck}
\ba
\tl{Z_{n}} \simeq \prod_{\a} \Biggl[ \int Dz_{\a}(x)
\exp \Biggl( -\fr{1}{2\D}\I z_{\a}^{2}(x) \Biggr) \Biggr] \\
\\
\times\sum_{\a_{1}...\a_{n}}^{K} \exp \Biggl[
\4 g\I \sum_{a} z^{2}_{\a_{a}}(x)- \2 g\I \sum_{a} z_{\a_{a}}(x)
[3\p_{a}(x)^2 -\Phi_{0}(x)^2 ] \Biggr] \ .
\ea
\ee

At this stage the problem (\ref{ccck}) seems to be similar to that
of the Random Energy Model (REM) \cite{rem}.
Leaving apart the details of the rigorous consideration, one can
obtain the correct solution for this problem in the framework of
the following simplified heuristic procedure.  Each term in the
exponent of the partition function (\ref{ccck}) is the sum
of $n$ values $\{z_{\a_{a}}\}$'s chosen out of the total
$K=2^{\kappa V}$ ones.  From the solution of the REM it
is seen that the leading contribution to the partition function
comes from the configurations in which in the summations
$\sum_{a=1}^{n}z_{\a_{a}}$ one takes $n/x_{0}$ different
$\a_{a}$'s, repeated $x_{0}$ times.  Here $x_{0}$ is a
parameter originally defined in the interval $1 \leq x_{0} \leq n$,
which (as usual in the replica formalism) turns into
$1 \geq x_{0} \geq 0$ in the limit $n \to 0$.  This parameter has
to be fixed by extremizing the free energy.  If such an extremum is
achieved for $x_{0}=1$, then one gets the RS solution, otherwise
the system appears to be in the 1step RSB state.

According to the above ansatz for the partition function
(\ref{ccck}) one gets:

\be
\lb{cccl}
\ba
\tl{Z_{n}} \simeq (2^{\kappa V})^{\fr{n}{x_{0}}} \prod_{c=1}^{n/x_{0}}
\Biggl[ \int Dz_{\a_{c}}(x) \exp \Biggl( -\fr{1}{2\D'}
\I z_{\a_{c}}^{2}(x) \\
\\
- \2 g\I z_{\a_{c}}(x) [3\sum_{b=1}^{x_{0}}\p_{cb}^{2}(x)
- x_{0}\Phi_{0}(x)] \Biggr) \Biggr] \ .
\ea
\ee
Here $(2^{\kappa V})^{\fr{n}{x_{0}}}$ is the combinatoric entropy
factor, and $\fr{1}{\D'} = \fr{1}{\D} - \2 g$.

After simple algebra one obtains:

\be
\lb{cccm}
\ba
\tl{Z_{n}} \simeq \exp \Biggl(
\fr{9}{8}g^{2}\D' \I \sum_{c=1}^{n/x_{0}}\sum_{b,b'=1}^{x_{0}}
\p_{cb}^{2}\p_{cb'}^{2} \\
\\
-\fr{3}{4}g^{2}\D' x_{0}\I \Phi_{0}\sum_{a=1}^{n} \p_{a}^{2}
+ \fr{1}{8}g^{2}\D' n x_{0}\I \Phi_{0}^{2}
+\fr{n}{x_{0}} \kappa V \ln 2 \Biggr) \ .
\ea
\ee
Coming back to the initial problem of integration over the
fluctuations $\p(x)$, Eq. (\ref{cccf}), one finds that the
second term in the
exponent (\ref{cccm}) gives an irrelevant shift of the
mass term in the Hamiltonian $H[\p(x)]$, while the first term
in (\ref{cccm}) provides the RSB structure of the matrix $g_{ab}$
in the interaction term:

\be
\lb{cccn}
\4\sum_{a,b=1}^{n}g_{ab}\p^{2}_{a}\p^{2}_{b}
\ee
The matrix $g_{ab}$ appears to have 1step RSB block structure
described by the parameters (in notations of the Section 2):

\be
\lb{ccco}
\ba
\tl{g} = g - u - \fr{9}{2}g^{2}\D'\\
\\
g_{1} = -u -\fr{9}{2}g^{2}\D'\\
\\
g_{0} = -u \ .
\ea
\ee

Since $g$, $u$, and $\D'$ are all positive by definition, one
finds the following
restrictions on the values of the interaction parameters:
$g_{0}<0$, $g_{1}<0$, $g_{1}\leq g_{0}$ and $\tl{g} > g_{1}$.

In the original REM problem \cite{rem}, after the partition
function is calculated, the parameter $x_{0}$ is fixed by
extremizing the resultant free energy with respect to
$x_{0}$. Here the parameter $x_{0}$ enters into the further
problem of integration over fluctuations which can be done
in terms of the RG procedure. However, in terms of the RG
technique one usually gets the results only for the singular
part of the free energy, which is actually small (in $\t$) compared
to the whole free energy. All that makes the problem of fixing
the parameter $x_{0}$ rather non-trivial.

It should be stressed, however, that aim of the considerations
of this Section was to identify the physical mechanism which may,
in principle, give rise to RSB perturbations whose treatment
then follows from the usual RG calculations of the critical
behaviour.  It will be very difficult to make our arguments
really precise.  It is clear, in view of the somewhat slippery
assumptions made above, that the actual contribution to the
interactions of the fluctuating fields, coming from the
non-perturbative degrees of freedom, could have an even
more sophisticated RSB structure, than the simple
1step form obtained above.  However, the
main point of the present discussion was to demonstrate
that, as far as the effective interactions of the fluctuating
fields are concerned, the RSB perturbations could exist in the
critical region.  To what extend such RSB perturbations are
relevant for the critical properties, can then be analysed
(as we have done) in terms of the traditional RG approach.
The results of Section 2 clearly demonstrate that whenever
the weak disorder is relevant (i. e. if the specific heat of
the pure system is positive), and if spontaneous RSB occurs,
then the the critical behavior is modified in a dramatic way.

\sectio{CONCLUSIONS}

We may summarize our conclusions concerning the random bond
$p$-component Heisenberg ferromagnet as follows.

\begin{itemize}
\begin{enumerate}

\item
The traditional fixed points\cite{new,newnew} in the weakly
random ferromagnet for the case when randomness is relevant are
only stable within the space of replica symmetric potentials.
Therefore, the corresponding results for the critical
exponents and other critical properties, are correct only as
long a replica symmetry breaking does NOT occur spontaneously.
When randomness is not relevant, our analysis reduces to the
standard one.\cite{new}

\item
Spontaneous replica symmetry breaking has a dramatic effect on
the renormalization group flows and therefore on the critical
properties.  At first order in $\epsilon$ and for $p$ not
close to 1 (the Ising limit), the stable fixed point
corresponds to one step replica symmetry breaking, in
close analogy with the random energy model.\cite{rem}
Presumably a calculation at higher order in $\epsilon$
would lead to predictions for how the correlation
functions would reflect a breaking of replica symmetry.

\item
At first order in $\epsilon$, there is an instability region
near $p=1$ where we find no stable fixed point in the
presence of replica symmetry breaking.  This result is
in close analogy to the known result in the absence of
replica symmetry breaking, namely that only at two-loop
order does one recover a stable fixed point\cite{newnew}
not present in first order.\cite{new}

\item
The replica symmetry broken fixed point is characterized by
a parameter $x_0$ whose value is not fixed in the order to which we
work here.  Accordingly, two main possibilities exist.
In the first case, the value of this parameter may be
determined in higher order in $\epsilon$, in which case
one would have universal exponents, amplitude ratios, etc.,
as usual for a critical point.  In the second case, one
would have continuously variable exponents and a corresponding
lack of universality.  We have no idea what aspects of the
randomness, when varied, would lead to variation in
the critical exponents.

\item
A key question, which remains unanswered, is whether or
not in the case of {\it arbitrarily} weak randomness
our model has spontaneous replica symmetry breaking.
We have given a scenario by which replica symmetry
can be spontaneously broken by interactions, $V_{\rm LP}$,
between the perturbative fluctuations (usually treated
within a renormalization group context) and fluctuations
about local mean field solutions (ignored in previous
treatments).  In this scenario we relate the partition
function due to fluctuations about local mean field
solutions to the random energy model.\cite {rem}
The crucial question, which we can not answer, is
whether $V_{\rm LP}$ is large enough that the
analogous random energy model is in its replica
symmetry broken phase or not.  So, it is possible
that there is a critical strength in the randomness,
below which replica symmetry breaking does not occur.

\item
One may mention related work.  For instance, the
existence of local solutions to the mean field
equations reminds one of the Griffith phase,
in which field derivatives of the free energy
are anomalous even outside the critical region \cite{griff}.
While the present paper was in preparation we learned
about similar RSB instability in the RG flows in the
2D random field XY model \cite{xy}.

\end{enumerate}
\end{itemize}

\vspace{5mm}

{\large \bf Acknowledgments}

VD acknowledges support from the International Association for
Promotion of Cooperation with Scientists from the Independent
States of the Former Soviet Union under contract No.
1010-CT93-0027(INATS) and ABH acknowledges support from
the Engineering and Physical Sciences Research Council
of the United Kingdom, Grant No. GR/K15633 (EPSRC).
Two of the authors (VD and ABH) thank the Theoretical
Physics Department of Oxford University for hospitality.

\pagebreak

\end{large}

\end{document}